\documentclass{aastex631}
\usepackage{subfigure}
\usepackage{makecell}
\usepackage{graphicx}	
\usepackage{multirow}
\usepackage{epstopdf}
\usepackage[normalem]{ulem}
\usepackage[figuresright]{rotating}
\usepackage{array}
\usepackage{tabularx}
\useunder{\uline}{\ul}{}

\makeatletter

\newcommand{\Rmnum}[1]{\expandafter\@slowromancap\romannumeral #1@}
\newcommand{\oiv}{O\,{\sc iv}}
\newcommand{\siiv}{Si\,{\sc iv}}
\newcommand{\kms}{km\,s$^{-1}$}
\makeatother
\shorttitle{Electron densities of transition region loops}
\shortauthors{liang et al.}

\begin{document}

\title{Electron densities of Transition Region Loops derived from IRIS \oiv\ spectral data}

\correspondingauthor{Zhenghua Huang}
\email{z.huang@sdu.edu.cn}

\author{Shiyu Liang}
\affiliation{Shandong Key Laboratory of Optical Astronomy and Solar-Terrestrial Environment, Institute of Space Sciences, Shandong University, Weihai, Shandong, 264209, China}
\affiliation{Key Laboratory of Solar Activity and Space Weather, National Space Science Center, Chinese Academy of Sciences, Beijing, 100190, China}

\author{Ziyuan Wang}
\affiliation{Shandong Key Laboratory of Optical Astronomy and Solar-Terrestrial Environment, Institute of Space Sciences, Shandong University, Weihai, Shandong, 264209, China}

\author[0000-0002-2358-5377]{Zhenghua Huang}
\affiliation{Shandong Key Laboratory of Optical Astronomy and Solar-Terrestrial Environment, Institute of Space Sciences, Shandong University, Weihai, Shandong, 264209, China}
\affiliation{Key Laboratory of Solar Activity and Space Weather, National Space Science Center, Chinese Academy of Sciences, Beijing, 100190, China}

\author[0000-0002-0210-6365]{Hengyuan Wei}
\affiliation{Shandong Key Laboratory of Optical Astronomy and Solar-Terrestrial Environment, Institute of Space Sciences, Shandong University, Weihai, Shandong, 264209, China}

\author[0000-0002-8827-9311]{Hui Fu}
\affiliation{Shandong Key Laboratory of Optical Astronomy and Solar-Terrestrial Environment, Institute of Space Sciences, Shandong University, Weihai, Shandong, 264209, China}

\author[0000-0001-9427-7366]{Ming Xiong}
\affiliation{Key Laboratory of Solar Activity and Space Weather, National Space Science Center, Chinese Academy of Sciences, Beijing, 100190, China}
\affiliation{College of Earth and Planetary Sciences, University of Chinese Academy of Sciences, Beijing, 100049, China}


\author[0000-0001-8938-1038]{Lidong Xia}
\affiliation{Shandong Key Laboratory of Optical Astronomy and Solar-Terrestrial Environment, Institute of Space Sciences, Shandong University, Weihai, Shandong, 264209, China}



\begin{abstract}
Loops are fundamental structures in the magnetized atmosphere of the sun.
Their physical properties are crucial for understanding the nature of the solar atmosphere.
Transition region loops are relatively dynamic and their physical properties have not yet been fully understood.
With spectral data of the line pair of \oiv\ 1399.8\,\AA\ \& 1401.2\,\AA\ ($T_{max}=1.4\times10^5$\,K) of 23 transition region loops obtained by IRIS, we carry out the first systematic analyses to their loop lengths ($L$), electron densities ($n_e$) and effective temperatures.
We found electron densities, loop lengths and effective temperatures of these loops are in the ranges of  $8.9\times10^{9}$--$3.5\times10^{11}$\,cm$^{-3}$, 8--30\,Mm and $1.9\times10^5$--$1.3\times10^6$\,K, respectively. 
At a significant level of 90\%, regression analyses show that the relationship between electron densities and loop lengths is $n_e[cm^{-3}]\varpropto (L[Mm])^{-0.78\pm0.42}$,
while the dependences of electron densities on effective temperatures and that on the line intensities are not obvious.
These observations demonstrate that transition region loops are significantly different than their coronal counterparts.
Further studies on the theoretical aspect based on the physical parameters obtained here are of significance for understanding the nature of transition region loops.
\end{abstract}


\section{Introduction}\label{sec:intro}

The magnetized solar upper atmosphere consists of various types of loops those are manifestations of plasma confined by magnetic field \citep{2003A&A...406.1089D,2015SSRv..188..211F}. 
Based on their temperatures, loops are normally classified into hot loops ($\geqslant$ 2 $\times$ 10$^{6}$ K), warm loops (1 - 2 $\times$ 10$^{6}$ K) and cool loops (10$^{5}$ - 10$^{6}$ K), which are normally associated with different dynamic natures and/or heating profiles.
Since loops are structures standing out from the ambient atmosphere, they can be well traced in the observations and their dynamics have been important objects for studying the nature of the mass and energy transportation in the solar atmosphere\,\citep{2014LRSP...11....4R}.

\par
Physical parameters of loops, such as velocity, density and temperature, are directly related to their pressure and radiative properties, and the distributions of these parameters along the length of a loop depend on the nature of heating therein. 
An empirical scaling law demonstrated by \citet{1978ApJ...220..643R} shows that maximum temperatures, pressures and lengths of hydrostatic coronal loops are subjected to a unique equation without any other free parameters.
Such a scaling law is in favor of coronal heating models involving coronal magnetic field\,\citep{1978ApJ...220..643R}.
Many more studies involving simulations and/or observations have confirmed that these physical parameters are crucial for unraveling the behaviours of heating and dynamics in loops\,\citep[see ][etc.]{1998Natur.393..545P,2006SoPh..234...41K,Patsourakos_2007,2008ApJ...677.1395W,2008ApJ...682.1351K,2013ApJ...771..115V,2017ApJS..229....4C,Polito_2018,2021ApJ...915...39H,2022A&A...664A..48C,2022A&A...667A.166C}.
By comparing the results of a 1D hydrodynamic simulation and IRIS observations,
\citet{Polito_2018} concluded that the response of loops in transition region emissions depends on the initial conditions, including temperature and density.

\par
Coronal loops, which have temperatures at an order of a million Kelvin, have been studied intensively since their discoveries by early rocket missions in the 1970s\,\citep{1973SoPh...33..265T}.
With data from the Extreme-Ultraviolet Imaging Telescope (EIT), \citet{2001ApJ...550.1036A} analyzed geometries, densities and temperatures of 30 coronal loops.
They concluded that steady state solutions for nonuniformly heated loops exhibit distinct characteristics compared to uniformly heated loops and that the pressure and density in nonuniformly heated loops are higher than those in uniformly heated loops by a factor up to two orders of magnitude.
By analyzing 67 loops with a wide range of apex temperatures and half-lengths observed by either the Transition Region and Coronal Explorer (TRACE) or the Soft X-Ray Telescope (SXT), \citet{2003ApJ...587..439W} found that static models with uniform heatings cannot explain the observed densities of both cool ($<$3\,MK) long loops and hot ($>$3\,MK) short loops.
The physical parameters of coronal loops have been studied in the era of SOHO, Hinode and STEREO.
\citet{2005A&A...439..351U} found that electron densities along a coronal loop decrease from $1.4\times10^9$\,cm$^{-3}$ at the footpoint to $0.9\times10^9$\,cm$^{-3}$ at the loop top.
Based on emission measurement method, \citet{2008ApJ...680.1477A} found electron densities of $(2.2\pm0.5)\times10^9$\,cm$^{-3}$ at the bases of seven coronal loops. 
For one active region loop observed by the EUV Imaging Spectrometer (EIS), \citet{2009ApJ...694.1256T} found from the region near its footpoint toward its apex the temperatures and electron densities vary from 0.8\,MK to 1.5\,MK and from 10$^{9}$\,cm$^{-3}$ to 10$^{8.5}$\,cm$^{-3}$ , respectively.
With EIS data, \citet{Xie_2017} derived densities, temperatures, filling factors, nonthermal velocities and magnetic strengths of 50 active region loops, and they found that most of the loops were in a state of overpressure compared to the prediction of \citet{1978ApJ...220..643R}, giving a constraint to a heating model.


\par
Transition region loops, which have temperatures from tens of thousands to hundreds of thousands of Kelvin, are usually difficult to identify owing to their dynamic nature, strong radiative background and line-of-sight (LOS) contamination from other features.
However, transition region loops with temperatures below 10$^{5}$ K can contribute significantly to the EUV output of the solar transition region\,\citep{1983ApJ...275..367F,1986SoPh..105...35D,1987ApJ...320..426F,1993ApJ...411..406D,1998ApJ...507..974F,2001ApJ...558..423F,2012A&A...537A.150S}.
In spectral observations of O\,{\sc v} ($2.5\times10^{5}$\,K) and O\,{\sc vi} ($3.2\times10^{5}$\,K) from the Solar Ultraviolet Measurements of the Emitted Radiation (SUMER) instrument and the Coronal Diagnostics Spectrometer (CDS) instrument onboard SOHO,
transition region loops could show line-of-sight velocities from a few tens to a hundred kilometers per second\,\citep{1997SoPh..175..511B,1998SoPh..182...73K,2000ApJ...533..535C,2003A&A...406..323D,2004A&A...427.1065T,2006A&A...452.1075D}.
High-resolution observations of loop structures in the H\,{\sc i} Ly$\alpha$ line taken with the Very High Angular Ultraviolet Telescope suggest that regions with temperatures within $1-3\times10^4$\,K and pressures within $1-3\times10^{-2}$\,Pa are suitable for the low-temperature ends of cool loops \citep{Patsourakos_2007,2010SoPh..261...53V}.
 
\par
Observations taken by the Interface Region Imaging Spectrometer\citep[IRIS,][]{ 2014SoPh..289.2733D} in the past decade provide an opportunity for in-depth studies to transition region loops\,\citep{2017RAA....17..110T,2019STP.....5b..58H,2021SoPh..296...84D}.
\citet{Huang_2015} reported on the first IRIS observations of a group of dynamic transition region loops with cross-sections ranging from 382 to 626 km.
They discovered clear evidence of siphon flows with velocities of 10--20\,\kms\ in those loops starting from the footpoints with significant magnetic cancellations, where impulsive heatings are suggestive to take place.
Transition region loops could be abundant in a flux-emerging region, where magnetic reconnection events frequently occur and magnetic geometry is reforming\,\citep{2014Sci...346C.315P,Huang_2015,2016ApJ...824...96T,2018ApJ...869..175H}.
Transition region loops might also form together with coronal loops in the same region, and that creates a complex mass and energy exchanging system in the solar atmosphere\,\citep{2018ApJ...869..175H,2019ApJ...887..221H,2022MNRAS.512.3149G}.
The extreme dynamic nature of transition region loops has also been confirmed by many other IRIS observations\,\citep{2017MNRAS.464.1753H,2019ApJ...874...56R,2021Symm...13.1390H}.

\par
Although the dynamic nature of transition region loops has been studied widely in the past decade, electron densities of transition region loops remain to be a missing piece of the puzzle.
The main reason is that the spectral lines suitable for density diagnostics in IRIS observations are too weak and there is a lack of suitable data.
In the present study, we focus on this issue and dig out a few IRIS datasets that allow us to perform good diagnostics to electron densities of transition region loops.
In the following, we describe the observations and data analyses in Section\,\ref{sec:obs}, present results and discussion in Section\,\ref{sec:res} and give conclusions in Section\,\ref{sec:con}.




\begin{table*}[]
\centering
\caption{Summaries of the datasets and corresponding loop numbers analyzed in the present study.}
\label{data}
\begin{tabular}{|c|c|c|c|c|c|c|c|c|}\hline
Data&
\multirow{2}{*}{Date and time} &
{FOV center} &
\multicolumn{3}{c|}{Raster} &
\multicolumn{2}{c|}{SJI} &
{Loop}
\\ \cline{4-8}
set& &(X,Y)&step size & exposure time & slit pixel & Cadence & pixel size
&Number \\ \hline
1&2013.12.27  19:23-19:57&316$^{\prime \prime}$,-268$^{\prime \prime}$ &0.35$^{\prime \prime}$ &4.0 s &0.167$^{\prime \prime}$ & 10 s &0.167$^{\prime \prime}$ & 1-6 \\ \hline
2&2013.12.27  21:02-21:36&324$^{\prime \prime}$,-270$^{\prime \prime}$ &0.35$^{\prime \prime}$ &4.0 s &0.167$^{\prime \prime}$ & 10 s &0.167$^{\prime \prime}$ & 7-11 \\ \hline
3&2013.12.27  22:38-23:11&338$^{\prime \prime}$,-267$^{\prime \prime}$ &0.35$^{\prime \prime}$ &4.0 s &0.167$^{\prime \prime}$ & 10 s &0.167$^{\prime \prime}$ & 12-13 \\ \hline
4&2014.07.01  19:55-20:57&18$^{\prime \prime}$,134$^{\prime \prime}$ &0.35$^{\prime \prime}$ &8.0 s &0.167$^{\prime \prime}$ & 39 s &0.167$^{\prime \prime}$ & 14-19 \\ \hline
5&2014.08.05  21:57-01:29&108$^{\prime \prime}$,-390$^{\prime \prime}$ &0.35$^{\prime \prime}$ &30.0 s &0.167$^{\prime \prime}$ & 132 s &0.167$^{\prime \prime}$ & 20-23 \\ \hline
\end{tabular}
\end{table*}


\begin{figure*}[!htb]
\centering
\includegraphics[width=\textwidth]{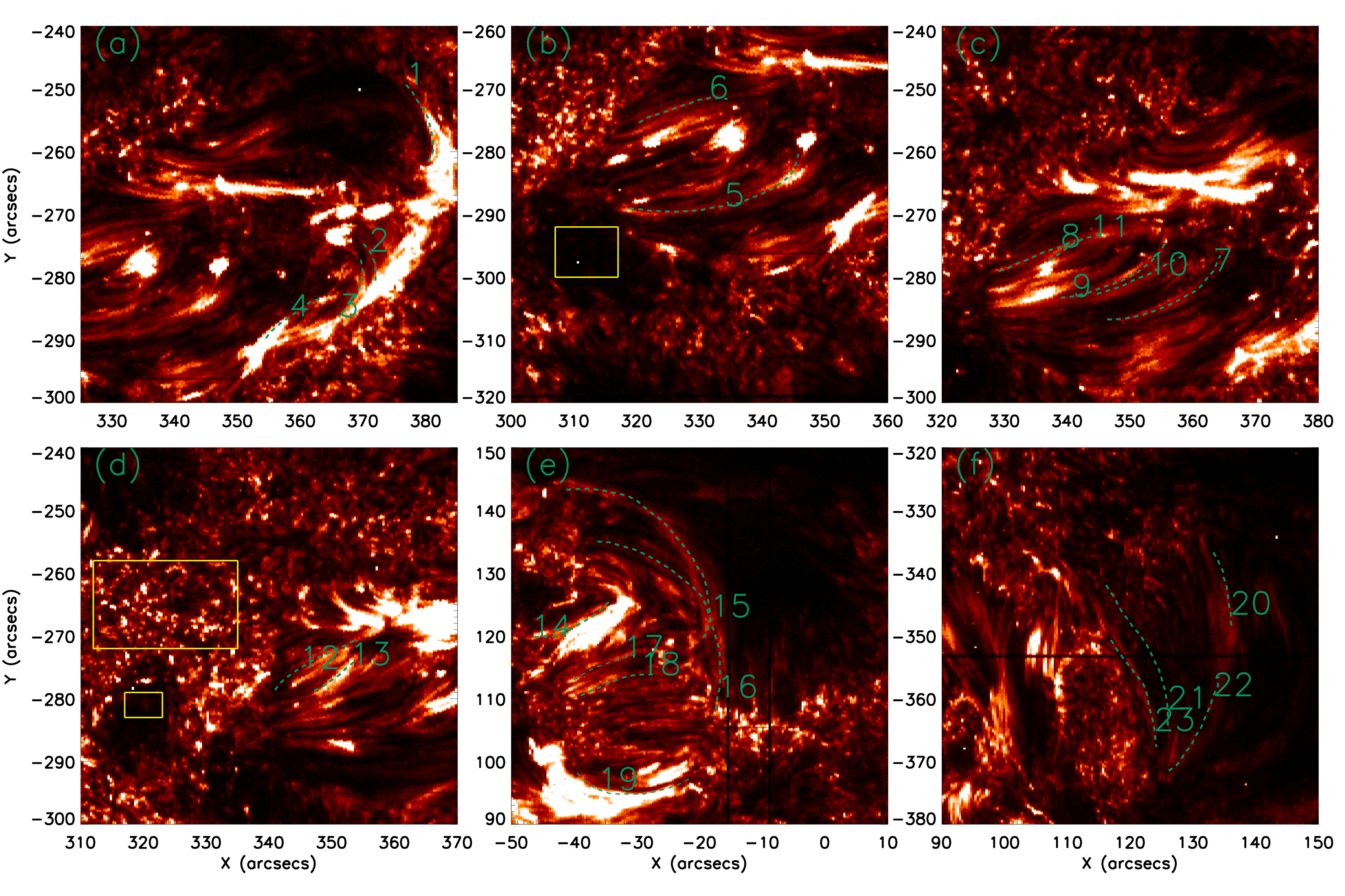}
\caption{Overviews of the transition region loops (green dashed lines denoted by numbers) selected for analyses as seen in the \siiv\,1402.8\,\AA\ raster images.  The yellow box in panel (b) denotes the area of the sunspot and those in panel (d) denote the areas of plage (the upper box) and pores (the lower box), which are determined by the \siiv\ emission and also the continuum around 2832\,\AA\ (images not shown).
}
\label{loops}
\end{figure*}

\begin{figure*}[!htb]
\centering
\includegraphics[width=\textwidth]{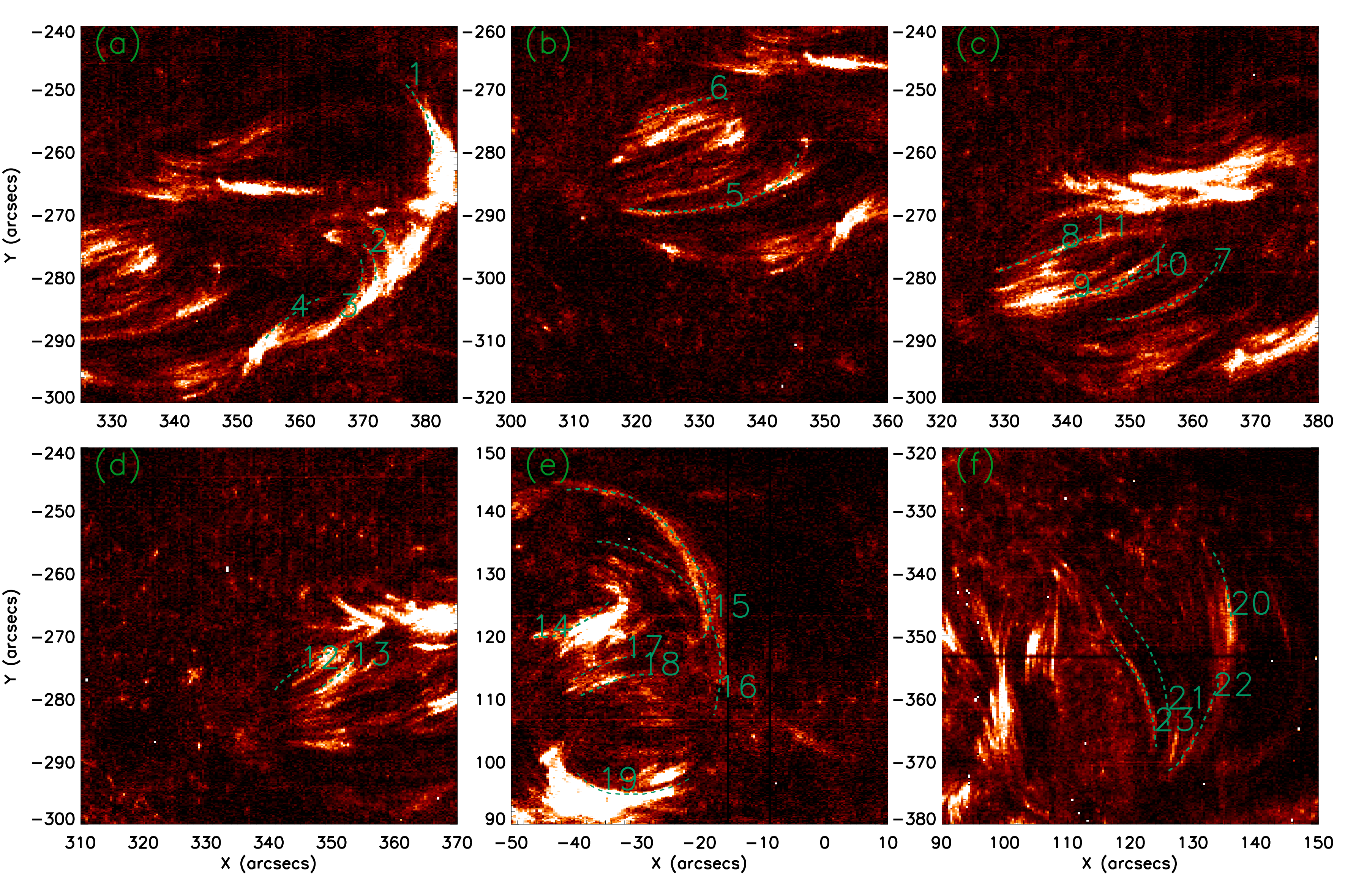}
\caption{Same as Figure\,\ref{loops}, but for the \oiv\,1401.2\,\AA\ raster images.
}
\label{loops_o4}
\end{figure*}

\section{Observations and data analyses} \label{sec:obs}
In this study, we analyze five IRIS datasets, which all are taken in the dense raster mode.
The details of the datasets are given in Table\,\ref{data}.
The first three datasets were targeting more or less the same region on the sun, but the selected loops are varied because the transition region loops evolve fast.
The IRIS data analyzed here are level-2 products, on which dark current removal, flat-field, geometrical distortion, orbital and thermal drift corrections have been applied.
In addition, we apply the radiometry calibration with the procedures described in the technical note ITN24 provided on the IRIS official website.

\par
Transition region loops are visually identified from the raster images of both \siiv\,1402.8 \AA\ and \oiv\ 1401.2\,\AA\ and the slit-jaw (SJ) images taken in the 1400\,\AA\, channel.
In Figure\,\ref{loops} and Figure\,\ref{loops_o4}, we show \siiv\,1402.8 \AA\ ($T_{max}=8\times10^4$\,K) and \oiv\ 1401.2\,\AA\ ($T_{max}=1.4\times10^5$\,K) raster images of the regions.
We can see that loops in these regions as seen in the two lines are well consistent, which is reasonable because both of the two lines are formed in the low transition region.
Because the difference in the formation temperatures of \siiv\ and \oiv\ is still significant, we select those loops that can be clearly seen in both raster images.
Thanks to the SJ data, we can trace the evolution of these loops and from that, we are able to locate their footpoints.
We would like to mention that transition region loops are very dynamic, and for each loop the spectrograph usually captures only a part of it rather than the full length.
Therefore, the full lengths of the loops shown as green dashed lines in Figures\,\ref{loops}\& \ref{loops_o4} are obtained from tracing their evolutions in the SJ data, and for each loop only the part captured by the spectrograph is used for the spectroscopic analyses.
To derive electron densities of the transition region loops, we use the line pair of \oiv\,1399.8\,\AA\ and 1401.2\,\AA\ that their line ratio is sensitive to electron density\,\citep{2015arXiv150905011Y}.
 Because the emissions of \oiv\,1399.8\,\AA\ lines are weak, not all transition region loops present in the forementioned raster images can be analyzed and only those having good signal-to-noise ratio in both \oiv\ lines are selected for further analyses.
The selected transition region loops (23 in total) are marked and denoted in Figure\,\ref{loops} and Figure\,\ref{loops_o4} and also listed in Table\,\ref{data} after the corresponding datasets.

\par
Because the background might provide a significant contribution to the total emissions, this contribution must be subtracted.
However, the background subtraction, which is normally achieved along the loop length, is not easy for the transition region loop here, because the ambient region is full of dynamic structures.
Here, for each dataset, we performed a background subtraction with the following steps: 
(1) remove pixels of spike, blank and fiducial marks from the \oiv\,1401.2\,\AA\ spectral data;
(2) extract the peaks of the \oiv\,\,1401.2\,\AA\ spectra for the rest of pixels;
(3) derive the mean ($\bar{I}$) and standard deviation ($\sigma$) of the \oiv\,1401.2\,\AA\ peaks;
(4) extract those pixels whose peak intensities fall within $\bar{I}\pm\sigma$;
(5) obtain the background spectrum by averaging those from the pixels given in step (4).
The obtained background spectrum is then subtracted from the loop spectra selected in the same dataset.
The emissions from these loops are found to be 3--40 times that of the background (see Table\,\ref{loopsdata}), indicating that they are much brighter structures than the ambient transition region.

\par
\begin{figure*}[!ht]
\centering
\includegraphics[width=\textwidth]{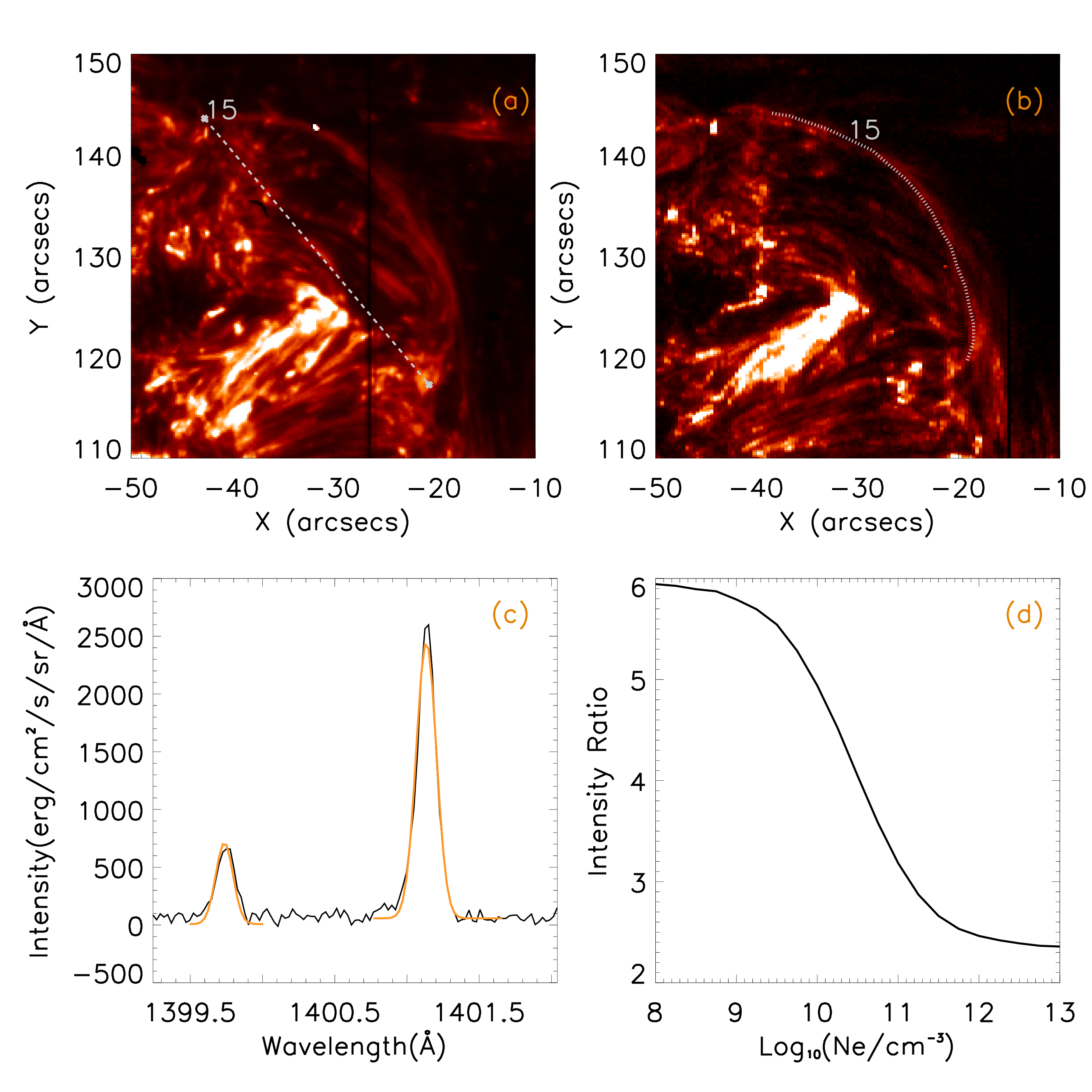}
\caption{An example showing the analyses to a transition region loop (No. 15). 
Overview of the loop in SJ 1400\,\AA\ image is shown in panel (a), on which we follow its evolution and determine its two footpoints (denoted by asterisks).
The distance between the two footpoints ($D$) is then obtained (white dashed line).
The overview of the loop region observed by spectrograph with \siiv\,1403\,\AA\ is shown in panel (b). 
Please bear in mind that the SJ image shown in panel (a) was taken at a single time while the image in panel (b) was taken in a raster mode, and thus many dynamic features appearing in (a) might not show in (b) and vice-versa.
The dotted line denotes the location of the loop, from which we produce the mean spectral profiles of \oiv\,1399.8\,\AA\ and 1401.2\,\AA\ as shown in panel (c).
In panel (c), the black line is the background-subtracted spectra, while the orange lines are the results of Gaussian fits to each of the \oiv\ spectral profiles.
The theoretical relation between electron densities and the line ratio of \oiv\,1401.2\,\AA\ to 1399.8\,\AA\ is given in panel (d).
The theoretical relation is obtained from version 10 of CHIANTI.
}
\label{loop15}
\end{figure*}

As an example, in Figure\,\ref{loop15} we show our analyses to loop No. 15.
The region containing the loop in SJ 1400\,\AA\ image is shown in panel (a).
Because only part of the loop might be bright in a certain snapshot, we have to determine its full apparent length ($L$) and footpoints by following its evolution in these SJ images (see an example in panel(a)).
The separation of the two footpoints ($D)$ of a loop is also achieved.
Because the \oiv\ emissions (especially the forbidden line at 1399.8\,\AA) are too weak, for most of the selected loops,  to have a good signal-to-noise ratio we have to produce the mean spectrum for each loop by averaging all pixels along its loop length.
In this way, although we ignore the variation of density along a loop, it should still have physical significance because transition region loops are dynamic and often filled with quasi-steady siphon flows\,\citep{Huang_2015} that facilitate material mixing along the loop and thus lead to nearly equal densities.
In one case that allows detailed analyses along the loop length, we confirm that densities along a transition region loop are more or less constant (see more details in Section\,\ref{sec:res}).
The pixels along a loop used for spectroscopic analyses are selected manually on the raster images (panel (b)) with the help from the SJ images, on which we can identify where and when the slit is scanning the loop.
Single Gaussian fits are then applied to the background-subtracted mean spectra to obtain the peak intensity ($I_p$), line centroid ($\lambda_0$) and line width ($w$) of each of the \oiv\ line profiles (panel (c)).
The line ratio of \oiv\ 1401.2\,\AA\ to 1399.8\,\AA\ is given as $r=(I_{p,1401.2}w_{1401.2})/(I_{p,1399.8}w_{1399.8})$.
The theoretical curve of line ratio vs. electron density ($n_e$) given by version 10 of CHIANTI database\,\citep{1997A&AS..125..149D,2021ApJ...909...38D} is shown in panel (d), from which we can derive the mean electron density of a transition region loop by using the observed line ratio as described above.

\par
Assuming that the line width results only from the instrumental broadening (0.026\,\AA) and thermal broadening, we also derive the effective temperatures of each transition region loop based on the spectral lines of \oiv\ 1401.2\,\AA\ and \siiv\,1402.8\,\AA\ ($T_{max}=8\times10^4$\,K), which are denoted as $T_{eff1}$ and $T_{eff2}$, respectively.
These effective temperatures represent the velocity distributions of the corresponding ions in the plasma\,\citep{1998ApJ...503..475T}.
For further discussion on the obtained effective temperatures, we calculate 
the equilibration time for ion and electron temperatures that is given by \citet{1962pfig.book.....S} and expressed as
$$
\tau_{eq}=\frac{158T_e^{3/2}}{n_eZ^2},
$$
where $\tau_{eq}$ has a unit of s, $T_e$ is electron temperature in a unit of Kelvin that is treated as the maximum formation temperature of the ion ($T_e=1.4\times10^5$ for \oiv), $n_e$ is the electron density in a unit of cm$^{-3}$, and $Z$ is ion charge that is 3 for \oiv.

\par
The \oiv\ 1399.8\,\AA\ might be blended by Fe\,{\sc ii}\,1399.96\,\AA\,\citep{2015arXiv150905011Y}.
The mean line profiles of all the loops have been checked manually.
We found the blending is negligible and they could be well fitted by single Gaussian functions.
To further confirm this result, the separation between the line centroids of \oiv\,1399.8\,\AA\ and 1401.2\,\AA\ ($\Delta\lambda_0$) is measured.
If the measured $\Delta\lambda_0$ is not too much away from the theoretical one (1.39\,\AA\ as given by CHIANTI), the blending of Fe\,{\sc ii}\,1399.96\,\AA\ is negligible (private communication with Dr. Peter Young).


\begin{table*}[]
\centering
\caption{Parameters derived from the \oiv\ spectral data for the studied transition region loops. $\Delta\lambda_0$ is the separation between the line centers of \oiv\ 1399.8\,\AA\ and 1401.2\,\AA. $I_{loop}/I_{back}$ is the ratio between the loop intensity and the background intensity of \oiv\,1401.2\,\AA. 
$n_e$ is electron density. 
$D$ is the distance between the two footpoints of a loop. 
$L$ is the length of a loop on the projected plane. 
 $T_{eff1}$ is effective temperature obtained from the \oiv\,1401.2\,\AA.
$T_{eff2}$ is effective temperature obtained from the \siiv\,1402.8\,\AA.
$\tau_{eq}$ is equilibration time for ion temperature and electron temperature.}
\label{loopsdata}
\begin{tabular}{|c|c|c|c|c|c|c|c|c|}\hline
Loop & $\Delta\lambda_0$ (\AA) & $I_{loop}/I_{back}$ & $n_e$ (cm $^{-3}$) & $D$ (Mm) & $L$ (Mm) & $T_{eff1}$ (K) & $T_{eff2}$ (K) & $\tau_{eq}$ (s) \\ \hline
1 & 1.38 & 17.97 & 7.6 $\times$ 10$^{10}$ & 8.64 & 9.33 & 6.4 $\times$ 10$^{5}$ & 8.8 $\times$ 10$^{5}$ & 1.2 $\times$ 10$^{-2}$ \\ \hline
2 & 1.40 & 9.52 & 5.2 $\times$ 10$^{10}$ & 10.21 & 12.16 & 4.0 $\times$ 10$^{5}$ & 5.2 $\times$ 10$^{5}$ & 1.8 $\times$ 10$^{-2}$ \\ \hline
3 & 1.39 & 12.51 & 1.0 $\times$ 10$^{11}$ & 8.42 & 9.14 & 3.2 $\times$ 10$^{5}$ & 5.4 $\times$ 10$^{5}$ & 9.0 $\times$ 10$^{-3}$ \\ \hline
4 & 1.38 & 7.18 & 9.0 $\times$ 10$^{10}$ & 9.69  & 9.90 & 9.4 $\times$ 10$^{5}$ & 1.1 $\times$ 10$^{6}$ & 1.0 $\times$ 10$^{-2}$ \\ \hline
5 & 1.39 & 7.26 & 1.3 $\times$ 10$^{11}$ & 20.04  & 21.50 & 4.2 $\times$ 10$^{5}$ & 5.1 $\times$ 10$^{5}$ & 7.0 $\times$ 10$^{-3}$ \\ \hline
6 & 1.39 & 7.25 & 3.5 $\times$ 10$^{10}$ & 11.20 & 11.28 & 2.0 $\times$ 10$^{5}$ & 4.2 $\times$ 10$^{5}$ & 2.6 $\times$ 10$^{-2}$ \\ \hline
7 & 1.38 & 6.01 & 1.1 $\times$ 10$^{11}$ & 19.50 & 20.77 & 1.3 $\times$ 10$^{6}$ & 2.3 $\times$ 10$^{6}$ & 8.0 $\times$ 10$^{-3}$ \\ \hline
8 & 1.41 & 8.40 & 6.8 $\times$ 10$^{10}$ & 10.02 & 10.04 & 2.7 $\times$ 10$^{5}$ & 3.3 $\times$ 10$^{5}$ & 1.4 $\times$ 10$^{-2}$ \\ \hline
9 & 1.38 & 8.61 & 5.7 $\times$ 10$^{10}$ & 20.23 & 20.83 & 3.8 $\times$ 10$^{5}$ & 5.8 $\times$ 10$^{5}$ & 1.6 $\times$ 10$^{-2}$ \\ \hline
10 & 1.41 & 7.09 & 3.0 $\times$ 10$^{10}$ & 13.15 & 13.32 & 6.7 $\times$ 10$^{5}$ & 5.3 $\times$ 10$^{5}$ & 3.1 $\times$ 10$^{-2}$ \\ \hline
11 & 1.39 & 8.57 & 6.6 $\times$ 10$^{10}$ & 7.93 & 8.03 & 2.5 $\times$ 10$^{5}$ & 4.7 $\times$ 10$^{5}$ & 1.4 $\times$ 10$^{-2}$ \\ \hline
12 & 1.39 & 7.84 & 5.1 $\times$ 10$^{10}$ & 11.71 & 11.80 & 4.9 $\times$ 10$^{5}$ & 6.8 $\times$ 10$^{5}$ & 1.8 $\times$ 10$^{-2}$  \\ \hline
13 & 1.38 & 13.03 & 6.1 $\times$ 10$^{10}$ & 8.89 & 9.02 & 2.9 $\times$ 10$^{5}$ & 4.5 $\times$ 10$^{5}$ & 1.5 $\times$ 10$^{-2}$ \\ \hline
14 & 1.39 & 15.25 & 3.5 $\times$ 10$^{11}$ & 10.00 & 10.42 & 6.8 $\times$ 10$^{5}$ & 1.3 $\times$ 10$^{6}$ & 3.0 $\times$ 10$^{-3}$ \\ \hline
15 & 1.40 & 13.38 & 3.5 $\times$ 10$^{10}$ & 24.12 & 29.64 & 4.2 $\times$ 10$^{5}$ & 6.2 $\times$ 10$^{5}$ & 2.7 $\times$ 10$^{-2}$ \\ \hline
16 & 1.39 & 13.34 & 1.1 $\times$ 10$^{11}$ & 22.15 & 25.09 & 5.2 $\times$ 10$^{5}$ & 8.4 $\times$ 10$^{5}$ & 8.0 $\times$ 10$^{-3}$ \\ \hline
17 & 1.38 & 13.54 & 1.0 $\times$ 10$^{11}$ & 9.14 & 9.16 & 2.7 $\times$ 10$^{5}$ & 5.6 $\times$ 10$^{5}$ & 9.0 $\times$ 10$^{-3}$ \\ \hline
18 & 1.40 & 14.71 & 1.0 $\times$ 10$^{11}$ & 9.34 & 9.42 & 4.7 $\times$ 10$^{5}$ & 8.0 $\times$ 10$^{5}$ & 9.0 $\times$ 10$^{-3}$ \\ \hline
19 & 1.38 & 43.89 & 3.1 $\times$ 10$^{10}$ & 14.93 & 17.46 & 6.9 $\times$ 10$^{5}$ & 1.3 $\times$ 10$^{6}$ & 3.1 $\times$ 10$^{-2}$ \\ \hline
20 & 1.38 & 6.41 & 1.0 $\times$ 10$^{10}$ & 21.29 & 22.16 & 2.9 $\times$ 10$^{5}$ & 3.8 $\times$ 10$^{5}$ & 9.4 $\times$ 10$^{-2}$ \\ \hline
21 & 1.38 & 3.51 & 2.9 $\times$ 10$^{10}$ & 18.70 & 19.00 & 1.9 $\times$ 10$^{5}$ & 2.2 $\times$ 10$^{5}$ & 3.2 $\times$ 10$^{-2}$ \\ \hline
22 & 1.38 & 3.63 & 8.9 $\times$ 10$^{9}$ & 15.60 & 16.14 & 5.7 $\times$ 10$^{5}$ & 5.7 $\times$ 10$^{5}$ & 1.0 $\times$ 10$^{-1}$ \\ \hline
23 & 1.39 & 3.12 & 1.1 $\times$ 10$^{11}$ & 9.13 & 9.71 & 3.5 $\times$ 10$^{5}$ & 5.8 $\times$ 10$^{5}$ & 8.0 $\times$ 10$^{-3}$ \\ \hline
\end{tabular}
\end{table*}

\begin{figure}[!ht]
\centering
\includegraphics[width=\textwidth]{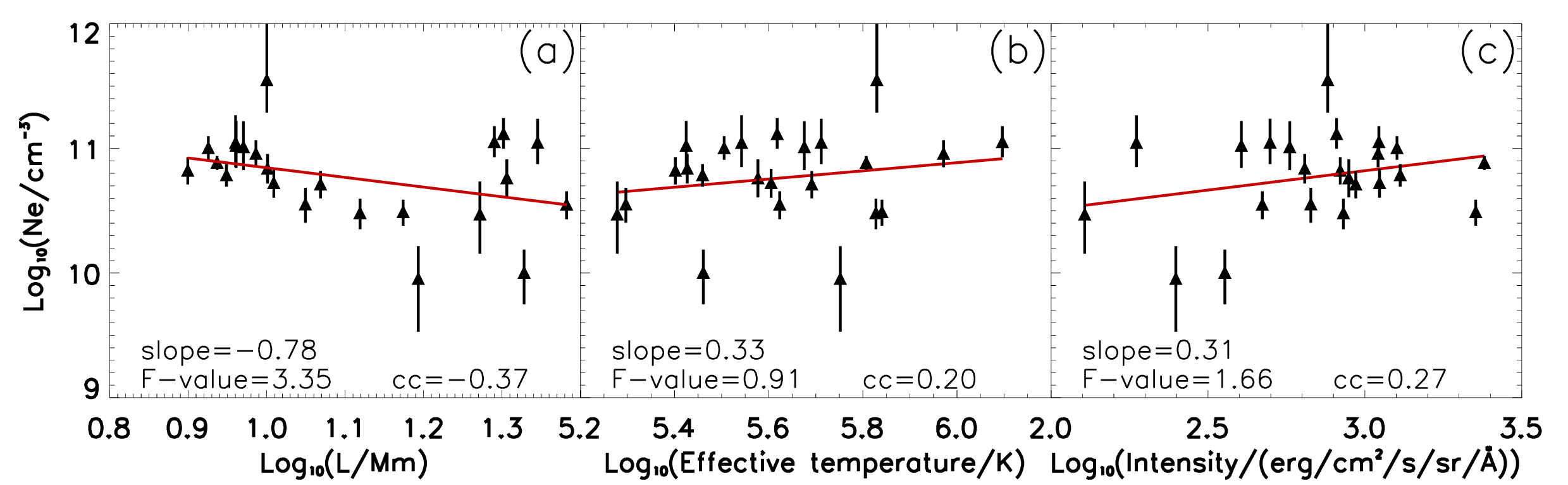}
\caption{Scatter plots of the studied transition region loops depicting the electron density {\it vs.} loop length (a), effective temperature obtained from the \oiv\,1401.2\,\AA\ (b) and the total fitted intensity of \oiv\,1401.2\,\AA\ (c).
The error bar of each data point is derived from the uncertainty given in the Gaussian fits of the \oiv\ profiles.
The parameters given by the linear regression analyses, including the slope, correlation coefficient (cc) and F-statistic (F-value), are also listed.
The red solid lines show the linear functions given by the regression analyses.} 
\label{len_den}
\end{figure}

\begin{figure}[!htb]
\centering
\includegraphics[width=\textwidth]{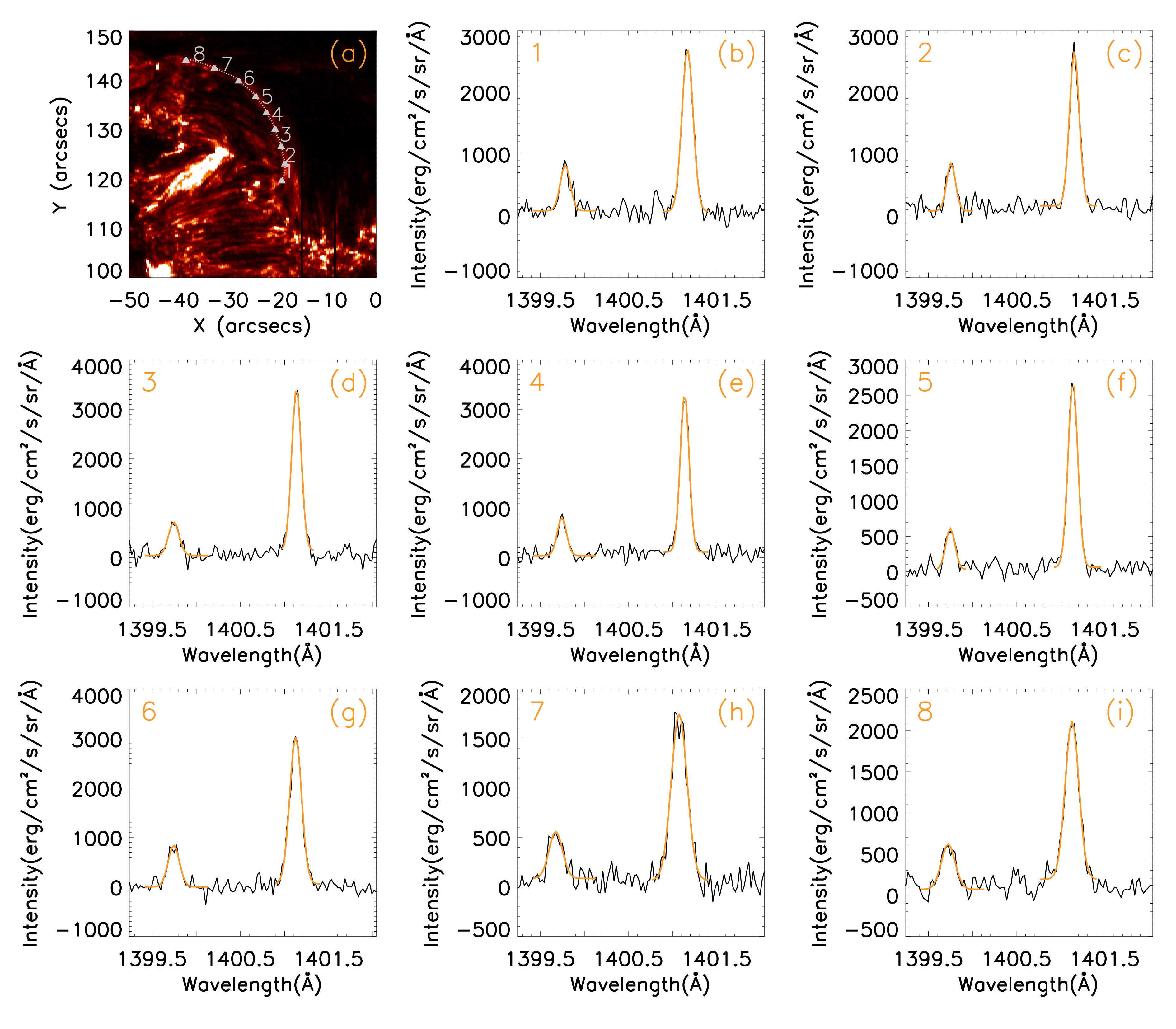}
\caption{(a): overview of loop No. 15 in IRIS \siiv\ raster image.
The full loop is divided into eight segments as denoted as 1--8.
The average spectra of these segments are shown as black lines in panels (b)--(i), on which the orange lines represent the Gaussian fits of \oiv\,1399.8\,\AA\, and 1401.2\,\AA.}
\label{loop15_gau}
\end{figure}

\begin{figure}[!htb]
\centering
\includegraphics[width=\textwidth]{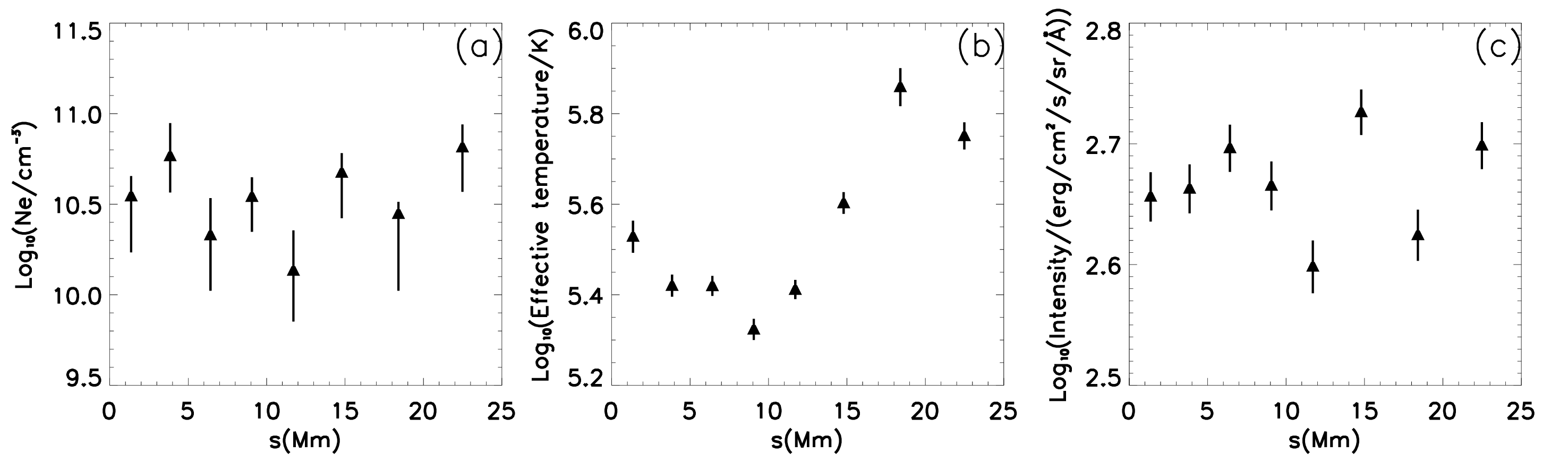}
\caption{The variations of electron densities (a), effective temperatures (b) and the total fitted intensities of \oiv\,1401.2\,\AA\ (c) along the loop No. 15.
The zero loop positions ($s=0$) start from the south footpoint of the loop.
The error bar of each data point is derived from the uncertainty given in the Gaussian fits of the \oiv\ profiles.}
\label{loop15_den_len}
\end{figure}

\section{Results and Discussion} \label{sec:res}



In Table\,\ref{loopsdata}, we list the parameters measured from 23 transition region loops, including the separation between \oiv\ 1399.8\,\AA\ and 1401.2\,\AA\ line centers ($\Delta\lambda_0$), the ratio between the loop intensity and the background intensity ($I_{loop}/I_{back}$), electron density ($n_e$), spatial separation between the two footpoints of a loop ($D$), the apparent loop length ($L$), effective temperatures ($T_{eff1}$ and $T_{eff2}$) and electron-ion equilibration time ($\tau_{eq}$) of each loop.
The values of $\Delta\lambda_0$ are close enough to the theoretical one (i.e. 1.39\,\AA), indicating that the emission of  Fe\,{\sc ii} 1399.96\,\AA\ is weak in these loops and its blending effect is negligible.
This also suggests that these loops might not have chromospheric counterparts at the same locations.

\par
The electron densities of these loops are in a range of $8.9\times10^{9}$--$3.5\times10^{11}$\,cm$^{-3}$.
The difference in electron density among these loops is more than an order of magnitude, showing that the transition region loops are a diverse group of structures in the solar atmosphere.
The lengths of these loops are in a range of 8--30\,Mm, which are similar to those of coronal bright points\,\citep{2012A&A...548A..62H} and much shorter than normal coronal loops\,\citep{2014LRSP...11....4R}.
The ratios of $D/L$ are in a range of 0.81--1, while most (19/23) of them are larger than 0.9.
These ratios could be affected by both the projection effect and/or the real geometries of these loops.

\par
The effective temperatures ($T_{eff1}$) of these loops derived from \oiv\ 1401.2\,\AA\ are in a range of $1.9\times10^{5}$--$1.3\times10^{6}$\,K.
These are significantly higher than the formation temperature of the ion.
While the effective temperature represents the velocity distribution of ions and the formation temperature represents that of the electrons,
the difference between the two indicates an imbalance between ions and electrons.
To assess whether there is a temperature imbalance, we calculate the equilibration time between ion temperature and electron temperature ($\tau_{eq}$) for each loop.
We found $\tau_{eq}$ for these loops are in a range of $3.0\times10^{-3}$--$1.0\times10^{-1}$\,s.

With inspections of SJ 1400\,\AA\ observations, we found that the variations of these loops have timescales of tens to hundreds of seconds.
These timescales are much larger than the equilibration times by a factor of a few orders of magnitude, and thus equilibrium between ions and electrons should be satisfied in these loops.
On the other hand, the exposure times of these observations are two to three orders of magnitude larger than the equilibration times and any transient nonequilibrium would not be resolved in the obtained spectra.
Therefore, while the observed effective temperatures generally exceed the formation temperature of the ion, it suggests that nonthermal motions (such as randomly oriented reconnection jets, waves, turbulence, etc.) are significant in the selected loops.

For a comparison, the effective temperatures obtained from \siiv\,1402.8\,\AA\ ($T_{eff2}$) are also listed in Table\,\ref{loopsdata}.
Overall, we found that the effective temperatures of \siiv\ are higher than those of \oiv.
$T_{eff2}$ is $1\sim2$ times of $T_{eff1}$.
This relation suggests that heating processes (described by nonthermal motions) at $8\times10^4$\,K are more significant than those at $1.4\times10^5$\,K.
These results should be taken into account for any further theoretical studies on the heatings of these loops.

\par
Figure\,\ref{len_den} (a), (b) and (c) show the distributions of these loops in the spaces of $log_{10}(n_e)$ vs. $log_{10}(L)$, $log_{10}(n_e)$ vs. $log_{10}(T_{eff1})$ and $log_{10}(n_e)$ vs. $log_{10}(I_{1401.2})$ ($I_{1401.2}$ is the total fitted intensity of \oiv\,1401.2\,\AA), respectively.
We found that correlation coefficients between these three pairs of parameters are 0.37, 0.20 and 0.27, which are relatively low.
In order to examine whether there is any trend in these distributions, we apply linear regression analyses and the obtained functions are drawn as red solid lines in Figure\,\ref{len_den}.
The F-statistic for the goodness-of-fit is then obtained, from which an F-test is then applied.
At a significant level of 90\%, an F-statistic has to be more than 2.961 to accept the obtained linear model.
From the results, we can tell there is a linear correlation between $log_{10}(n_e)$ and $ log_{10}(L)$, and the regression analysis shows a relation of $log_{10}(n_e)\varpropto -0.78 log_{10}(L)$.
The linear coefficient has a one-sigma error of 0.42.
For the relations of  $log_{10}(n_e)$ vs. $T_{eff1}$ and $log_{10}(n_e)$ vs. $log_{10}(I_{1401.2})$,
the obtained regression models are unlikely to be accepted.
%

\par
The relation of $log_{10}(n_e)$ vs. $log_{10}(L)$ is sensitive to heating status of loops\,\citep{1978ApJ...220..643R,2011ApJ...736....3V,2014PASJ...66S...7B,2019ApJ...880...80B}.
Based on the assumption of energy balance among heating, conduction and radiation in hydrodynamics static loops, \citet{1978ApJ...220..643R} derived that maximum temperature ($T_{max}$), pressure ($p$) and size ($L$) of coronal loops satisfy $T_{max}\sim 1.4\times 10^3 (pL)^{1/3}$.
Assuming that $T_{max}$ is the same for all loops of a group, one would obtain $n_e \varpropto L^{-1}$.
By analyzing 30 active region loops based on a novel dynamic stereoscopy method, \citet{Aschwanden_1999} found that pressures and lengths of the loops satisfy $p\varpropto L^{-0.41 \pm 0.12}$.
With a 3D MHD model, \citet{2014PASJ...66S...7B} successfully reproduce coronal loops similar to observations, and from the output of their numerical experiments they found electron densities and lengths of coronal loops have a relation of $n_e\varpropto L^{-3/7}$.
The exponent found here ($-0.78\pm0.42$) is significantly different from those of coronal loops reported previously.

This indicates that the transition region loops and the ways they are being heated might differ from their coronal counterparts.
Studies have also shown that such a relation could be fundamentally different in coronal loops while considering different mechanisms of thermal conduction\,\citep{2019ApJ...880...80B}.
The dynamic nature in transition region loops as studied here can easily introduce differences in the ways of conduction and thus lead to a relationship as found.
Moreover, the density of a transition region loop could be affected by many other factors under a dynamic condition, and it is reasonable that they do not follow hydrostatic law.




\par
The relation between electron densities and effective temperatures is not significant, suggesting that the electron density of a transition region loop is not strongly depended on the heatings.
Such a relation could be rational because transition region loops are relatively dynamic and thus their densities are affected by many other factors, such as the sources of siphon flows, the stages of their emergence, the influence of the surrounding features, etc.
We also calculate the effective temperatures from the background spectra (see descriptions on background subtraction for the definition) and three typical areas of the ambient transition region, including a plage, a pore and a sunspot (see Figure\,\ref{loops}).
The $T_{eff1}$ of the background spectra is about $5\times10^5$\,K, and those for plage, pore and sunspot are $4.9\times10^5$\,K, $2\times10^5$\,K and $3.2\times10^5$\,K, respectively.
The effective temperatures of many transition region loops are not obviously different than these typical regions.

\par
The weak correlation between electron densities and line intensities may be attributed to the facts that emissions of the spectral line in the transition region are affected by many complex processes.
For example, collisional processes (ionisation/recombination and excitation/de-excitation), photoionization and double electron recombination all have complex dependence on density.
Any of these processes might actually work in these transition region loops.


\par
One case, No. 15 loop, is bright enough and we can carry out electron density diagnoses for subsections along its length.
In Figure\,\ref{loop15_gau}, we show the overview of the loop that is divided into eight segments (panel (a)) and the mean spectra from these segments (panels (b)--(i)).
The variations of electron densities, effective temperature and \oiv\ intensities along the loop are shown in Figure\,\ref{loop15_den_len}.
One can see that there is no obvious trend in the variation of densities and intensities (Figure\,\ref{loop15_den_len}(a)\&(c)).
This relatively constant electron density along this loop confirms that the averaged density could be representative of those along a transition region loop.
More studies are certainly needed in the future, especially the coming EUVST and MUSE might shed more light on this issue using other spectral line-pairs with stronger emissions.
In contrast, the variation of effective temperature shows clearly that effective temperature is decreasing from the footpoints to loop top.
This demonstrates that heating of this loop concentrates at its footpoints, being consistent with the IRIS SJ observations.

\section{Conclusions} \label{sec:con}
Loops are fundamental structures of the solar atmosphere.
Their physical properties are sensitive to heating processes therein and thus are crucial for understanding their natures.
Transition region loops are an important class of loops, but so far there is not any systematic study on their physical parameters, especially the electron density.
In this paper, using IRIS spectral data of \oiv\ 1399.8\,\AA\ and 1401.2\,\AA\ we report on diagnoses of electron densities, effective temperatures, loop lengths and emissions to 23 transition region loops.
Furthermore, we analyze the relations between these parameters that can be used for comparison with potential numerical experiments in the future.

\par
With these observations,
the projected lengths ($L$) of these loops are found to be in the range of 8--30\,Mm.
While the distance between two footpoints ($D$) of each loop is also measured,
we found that the ratios of $D/L$ are in the range of 0.81--1.
The electron densities ($n_e$) of these loops range from $8.9\times10^{9}$\,cm$^{-3}$ to $3.5\times10^{11}$\,cm$^{-3}$, and the effective temperature obtained from \oiv\ ($T_{eff1}$) are in the range of $1.9\times10^{5}$--$1.3\times10^{6}$\,K those are significantly larger than the formation temperature of the ion.
The equilibration times between the ion temperature and electron temperature of these loops are in the range of $3.0\times10^{-3}$--$1.0\times10^{-1}$\,s, much smaller than the timescales of the loop evolution and also the exposure times by factors of a few orders of magnitude, giving evidence that the two temperatures should be in balance in these loops.
Therefore, the nonthermal motions in these loops should be significant and they account for the exceeded effective temperatures.
We also found that the effective temperatures obtained from \siiv\ are overall larger than those from \oiv, suggesting that in these loops nonthermal motions at $8\times10^4$\,K are more significant than those at $1.4\times10^5$\,K.

\par
The relations among electron densities ($n_e$), loop lengths ($L$), effective temperature and the total line intensity of \oiv\,1401.2\,\AA\ are analyzed.
By applying regression analyses, at a 90\% significant level we can tell that the relationship between electron densities and loop lengths is $n_e\varpropto L^{-0.78\pm0.42}$, where $n_e$ and $L$ have units of cm$^{-3}$ and Mm, respectively.
The analyses also show that the dependences of electron densities on effective temperatures and on line intensities are not obvious.

\par
In one transition region loop that allows detailed analyses, we found that the variation of electron densities along the loop length is not obvious while effective temperatures at the footpoints are clearly larger than those near the loop apex.
These results suggest that heating to this loop concentrates on its footpoints while the heated plasma can be redistributed along the loop very efficiently.

\par
These observations suggest that transition region loops are significantly different than their coronal counterparts.
The physical parameters obtained here should be a good start for further numerical experiments to understand the natures of many interesting patterns in transition region loops, such as siphon flows\,\citep{Huang_2015} and decayless oscillations\,\citep{2024A&A...681L...4G}.

\begin{acknowledgments}
We are grateful to the anonymous referee for the constructive comments and suggestions that help improve a lot the paper.
We are grateful to Dr. Peter Young for helping understand the blendings in the \oiv\ lines.
We thank Dr. Kai Wang for the fruitful discussion.
This research is supported by the National Natural Science Foundation of China (42230203, 42174201, 41974201, 42074208) and National Key R\&D Program of China No. 2021YFA0718600.
Project Supported by the Specialized Research Fund for State Key Laboratories.
IRIS is a NASA small explorer mission developed and operated by LMSAL with mission operations executed at NASA Ames Research Center and major contributions to downlink communications funded by ESA and the Norwegian Space Centre. 
CHIANTI is a collaborative project involving the University of Cambridge (UK), the NASA Goddard Space Flight Center (USA), the George Mason University (GMU, USA) and the University of Michigan (USA).
\end{acknowledgments}

\bibliography{references}{}
\bibliographystyle{aasjournal}
\end{document}